\documentclass{jpp}
\usepackage{subeqn}
\usepackage{epsfig}

\let\realverbatim\verbatim
\let\realendverbatim\endverbatim
\renewenvironment{verbatim}
  {\par\addvspace{6pt plus 1pt minus 2pt}\realverbatim}
  {\realendverbatim\addvspace{6pt plus 1pt minus 2pt}}

\newcommand\dynpercm{\nobreak\mbox{$\;$dynes\,cm$^{-1}$}}

\newcommand\Real{\mbox{Re}}     
\newcommand\Imag{\mbox{Im}}     
\newcommand\Rey{\mbox{\textit{Re}}}  
\newcommand\Pran{\mbox{\textit{Pr}}} 
\newcommand\Ai{\mbox{\textit{Ai}}}
\newcommand\Bi{\mbox{\textit{Bi}}}

\ifprodtf \else
  \checkfont{eurm10}
  \iffontfound
    \IfFileExists{upmath.sty}
      {\typeout{^^JFound AMS Euler Roman fonts on the system,
                   using the 'upmath' package.^^J}%
       \usepackage{upmath}}
      {\typeout{^^JFound AMS Euler Roman fonts on the system, but you
                   don't seem to have the}%
       \typeout{'upmath' package installed. JPP.cls can take advantage
                 of these fonts,^^Jif you use 'upmath' package.^^J}%
       \providecommand\mathup[1]{##1}%
       \providecommand\mathbup[1]{##1}%
       \providecommand\upi{\pi}%
       \providecommand\umu{\umu}%
      }
  \else
    \providecommand\mathup[1]{#1}%
    \providecommand\mathbup[1]{#1}%
    \providecommand\upi{\pi}%
    \providecommand\umu{\mu}%
  \fi
\fi

\ifprodtf \else
  \checkfont{msam10}
  \iffontfound
    \IfFileExists{amssymb.sty}
      {\typeout{^^JFound AMS Symbol fonts on the system, using the
                'amssymb' package.^^J}%
       \usepackage{amssymb}%

      }{}
  \else
  \fi
\fi

\ifprodtf \else
  \IfFileExists{amsbsy.sty}
    {\typeout{^^JFound the 'amsbsy' package on the system, using it.^^J}%
     \usepackage{amsbsy}}
    {\providecommand\boldsymbol[1]{\mbox{\boldmath $##1$}}}
\fi

\newcommand\ssC{\mathsf{C}}    
\newcommand\sfsP{\mathsfi{P}}  

\newcommand\bmath[1]{\ensuremath{\boldsymbol{#1}}}

\newcommand{\be}{\begin{equation}}
\newcommand{\ee}{\end{equation}}

\newcommand{\eqa}{\begin{eqnarray*}}
\newcommand{\eqe}{\end{eqnarray*}}
\newcommand{\eqnu}{\begin{eqnarray}}
\newcommand{\eqne}{\end{eqnarray}}

\newcommand{\eeq}{\end{eqnarray}}

\newdefinition{definition}[theorem]{Definition}

\title[Journal of Plasma Physics]
{Minimum Dissipative Relaxed States Applied to Laboratory and
Space Plasmas}

\author[Dastgupta et al]
{B. \ls D\ls A\ls S\ls G\ls U\ls P\ls T\ls A\ls , D\ls A\ls S\ls T\ls G\ls E\ls
E\ls R\ns S\ls H\ls A\ls I\ls K\ls H\ls, Q. \ls H\ls U\ls and G.\ls P.\ls Z\ls
A\ls N\ls K\ls}
\affiliation{Institute of Geophysics and Planetary Physics (IGPP),\\
University of California, Riverside, CA 92521. USA.\\
{\tt Email:dastgeer@ucr.edu}}
\date{April 17 2008, Revised May 6, 2008, Accepted May 7, 2008}
\pubyear{2008} 
\volume{00}
\part{0}
\pagerange{\pageref{firstpage}--\pageref{lastpage}}
\doi{S0963548301004989}

\begin{document}

\label{firstpage}
\maketitle

\begin{abstract}
The usual theory of plasma relaxation, based on the selective
decay of magnetic energy over the (global) magnetic helicity,
predicts a force-free state for a plasma. Such a force-free state
is  inadequate to describe most  realistic plasma systems
occurring in laboratory and  space plasmas as it produces a zero
pressure gradient and  cannot couple magnetic fields with flow. A
different theory of relaxation has been proposed by many authors,
based on a well-known principle of irreversible thermodynamics,
the principle of minimum entropy production rate which is
equivalent to the minimum dissipation rate (MDR) of energy. We
demonstrate the applicability of minimum dissipative relaxed
states to various self-organized systems of magnetically confined
plasma in the laboratory and in the astrophysical context. Such
relaxed states are shown to produce a number of basic
characteristics of laboratory plasma confinement systems and solar
arcade structure.
\end{abstract}


\section{Introduction and Motivation:}

Self-organization is a natural process [Ortolani and Schnack,
1993] in which a continuous system evolves toward some preferred
states showing a form of order on long scales. Examples of such
self-organization processes are ubiquitous in nature and in
systems studied in the laboratory. These ordered states are very
often remarkably robust, and their detailed structures are mostly
independent of the way the system is prepared.  The final state to
which a system evolves generally depends on the boundary
conditions, inherent geometry of the particular device, but is
relatively independent of the initial conditions.
Self-organization in plasma is generally called Plasma Relaxation.
The seminal theory  for plasma relaxation first proposed by Taylor
(1974) yields a force-free state with zero pressure gradient.
Taylor's theory was the first to explain the experimentally
observed field reversal of the Reversed Field Pinch (RFP).

Self-organized states with zero pressure gradient are often far
from experimentally realizable scenarios. Any realistic magnetic
configuration confining plasma must have a non-zero pressure
gradient. Extensive numerical simulations [ Sato et al., 1996, Zhu
et al., 1995] have established the existence of self-organized
states of plasma with finite pressure, i.e., these states  are
observed to be governed by the magnetohydrodynamic  force balance
equation. Numerical simulations [Dasgupta et al, 1995, Watanabe et
al, 1997] and experiments [Ono et al., 1993 ] have established
that counterhelicity merging of two spheromaks can produce a
Field-Reversed Configuration (FRC). The  FRC has a zero toroidal
magnetic field, and the plasma is confined entirely by poloidal
magnetic field. It has a finite pressure,  nonzero perpendicular
component of current  and is more often characterized by a high
value of plasma beta. The stability and long life categorize FRC
as a relaxed state, which is not obtainable from Taylor model of
relaxation. Since spheromaks  are depicted as a Taylor state
[Rosenbluth and Bussac, 1979],  formation of a FRC from the
merging of two counterhelicity spheromaks is a unique process from
the point of view of plasma relaxation,  where  a non-force free
state (hence non-Taylor state)  emerges from the coalescence of
two Taylor states. In  astrophysical context, particularly solar
physics scenario, a 3-D MHD simulation by Amari and Luciano
[2000], among others, showed that, after the initial helicity
drive, the final ``relaxed state is far from the ... linear
force-free model that would be predicted by Taylor's conjecture"
and they suggested to derive an alternative variational approach.
Such  physical processes call for an alternate model for plasma
relaxation to broadly accommodate a large number of  these
observations.

In this context, an important work by Turner deserves special
attention. Turner first showed (1986) that a relaxed magnetic field
configuration which can support/confine a plasma with a finite
pressure gradient. Adopting a two-fluid approach, a model of
magnetofluid relaxation is constructed for Hall MHD, under the
assumptions of minimum energy (magnetic plus fluid) with the
constraints of global magnetic helicity hybrid helicity, axial
magnetic flux and fluid vorticity. Euler-Lagrange equations resulted
from such variational approach show the coupling between the magnetic
field and the fluid vorticity.  The solutions for such coupled
equations are shown to be achieved by the linear superposition of the
eigen-vectors of the 'curl' operator (force-free states). These
solutions yield magnetic field configurations which can confine plasma
with a finite pressure gradient.

The principle of  ``minimum rate of entropy production" formulated
by Prigogine and others [Prigogine, 1947]  is believed to play a
major role in many problems of irreversible thermodynamics. This
principle states that for any steady irreversible processes, the
rate of entropy production is minimum. For most irreversible
processes  in nature, the minimum rate of entropy production is
equivalent to the minimum rate of energy dissipation.  A
magnetized plasma is such a dissipative system and it is
appropriate to expect that  the principle of minimum dissipation
rate of energy has a major role to play in a magnetically confined
plasma.   It is worth mentioning that Rayleigh [1873] first used
the term ``principle of least dissipation of energy" in his works
on the propagation of elastic waves in matter.  Chandasekhar and
Woltjer [1958]  considered a relaxed state of plasma with minimum
Ohmic dissipation with  the constraint of constant magnetic energy
and obtained an equation for the magnetic field involving a
`double curl' for the magnetic field and remarked that the
solution of such equation is ``much wider" than the usual force
free solution.

Montgomery and Phillips  [1988] first used the principle of
minimum dissipation rate  (MDR)  of energy in an MHD problem to
understand the steady the steady profile of a RFP under the
constraint of a constant rate of supply and dissipation of
helicity with the boundary conditions for a conducting wall.
Farengo et al [Farengo et al, 1994,1995, 2002, Bouzat, 2006] in a
series of papers applied  the principle of MDR to a variety of
problems ranging from flux-core spheromak to tokamak.

In this work, we present an alternative scenario for plasma
relaxation which is based on the principle of minimum dissipation
rate (MDR) rate of energy. This model gives us a non-force free
magnetic field, capable of supporting finite pressure gradient. We
demonstrate the this model can reproduce some of the basic
characteristics of most of the plasma confining devices in the
laboratory  as well as the arcade structure of solar magnetic
field. A two-fluid generalization of this model for an open system
couples magnetic field with flow - so this model is applicable to
other astrophysical situations.

The plan of the paper is as follows: In section 2, we present a
single fluid MDR model for closed system and briefly discuss its
applicarions to some of the laboratory plasma confinement devices,
such as RFP, Spheromak, Tokamak and FRC. We present the result of
a recent numerical simulation to justify our choice of the energy
dissipation rate as an effective minimizer in our variation
problem. In section 3 we describe a generalized version of the MDR
model for open two fluid system and its application to solar
arcade problem. We summarize a numerical extrapolation method for
non-force free coronal magnetic field based on our MDR model.
Section 4 concludes our paper.
\section{Single fluid MDR based model}

To obtain both the field reversal and a finite pressure gradient
for RFP, Dasgupta et al.[ Dasgupta et al., 1998] considered the
relaxation of a slightly resistive and turbulent plasma using the
principle of MDR under the constraint of constant global helicity.
With the global helicity $K_M$ and the dissipation rate $R$
defined as,
\begin{equation}
K_M = \int_V {\bf A}\cdot{\bf B}dV; \qquad R = \int_V \eta {\bf
j}^2 dV, \end{equation}

the variational principle $\delta (R-\Lambda K)$ leads to the
following Euler-Lagrange equation,

\begin{equation}\nabla\times\nabla\times\nabla\times{\bf B}=
\Lambda{\bf B}.\end{equation}

 Solutions of the above equation can be constructed as a
 superposition of the force free equation using
 Chandrasekhar-Kendall (CK)  eigenfunctions [Chandraskhar and
 Kendall, 1957]

\begin{equation}B =  \alpha_1 B_1 + \alpha_2B_2  + \alpha_3B_3,
\end{equation}
where $\alpha_1, \alpha_2, \alpha_3 $ are constants, to be
determined from the boundary conditions, and ${\bf B}_i's$ are
obtained from \begin{equation} \nabla\times{\bf B}_i = \lambda_i
B_i;\quad    (i = 1,2,3);\quad \lambda_i^3 =
\Lambda.\end{equation} We have used the analytic continuation of
CK eigenfunctions in the complex  domain.  However, one can easily
ascertain that the resulting magnetic field B is real. Moreover,
since the superposition of  force free fields with different
eigenvalues $\lambda_i$ is not a force free field, the resultant
magnetic field B is not force free, ${\bf J}\times {\bf B}\neq 0
$.  An explicit form of the solution in cylindrical geometry with
the boundary conditions for an RFP (perfectly conducting wall) has
shown two remarkable features: (1) Such a state can support a
pressure gradient; and (2) Field reversal is found in states that
are not force free.

\begin{figure}[ht]
\epsfig{file=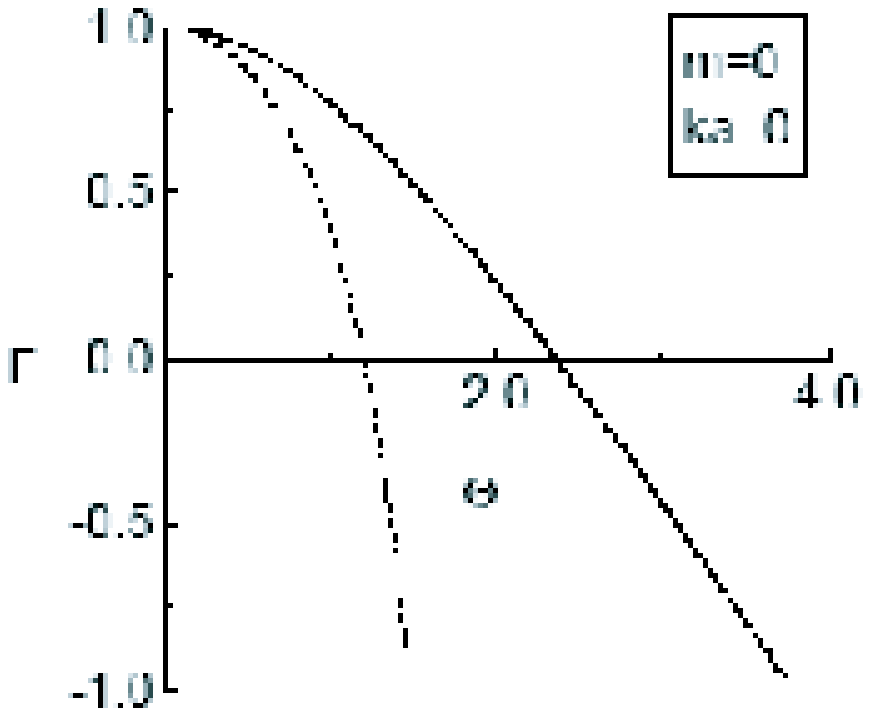, width=10.cm, height=8.cm}
\epsfig{file=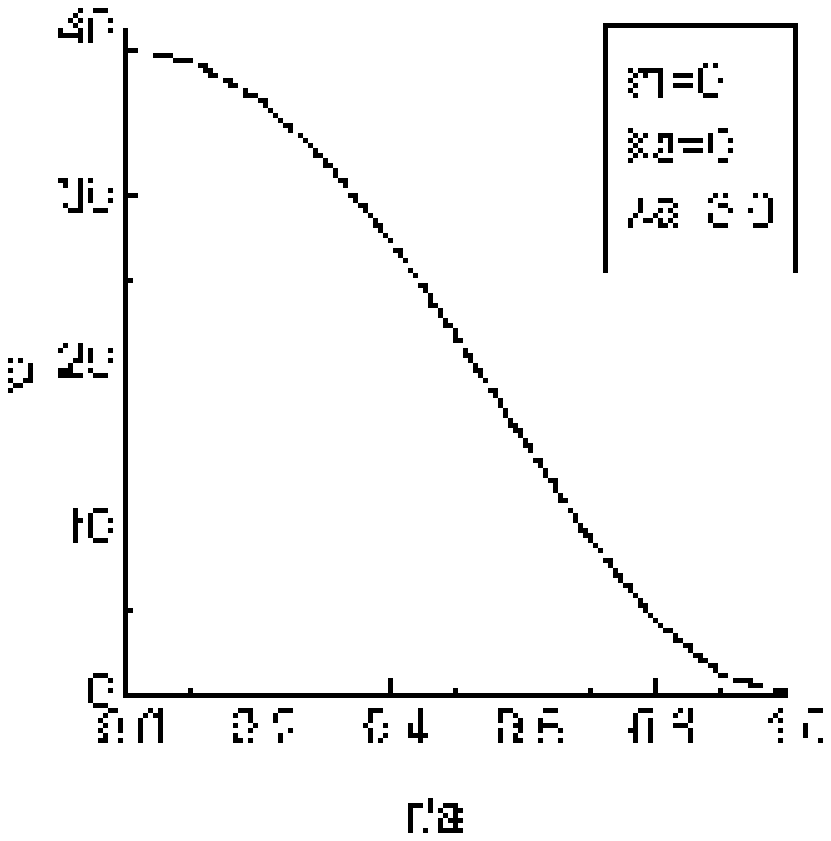, width=10.cm, height=8.cm}
\caption{\small\label{fig:RFPI} Left panel: Plot of $F =
B_z(a)/\left<B_z\right> $ and $\Theta =
B_\theta(a)/\left<B_z\right> $ ($<..> $ is the volume average and
$ a$ is the radial distance) showing the field reversal of RFP
obtained from MDR model. The dotted curve represents the plot for
the minimum energy state of Taylor. Right Panel: Pressure profile
of RFP obtained from MDR model.(\textit{from Dasgupta et al.,
1998}).}
\end{figure}

 In Figure \ref{fig:RFPI} a  plot of the field reversal
parameter $F$ against the pinch parameter $\Theta$ is shown. It is
found that axial field reverses at a value $\Theta \simeq 2.2$;
and the pressure profile of the RFP obtained from our MDR model.

\begin{figure}[ht]
\epsfig{file=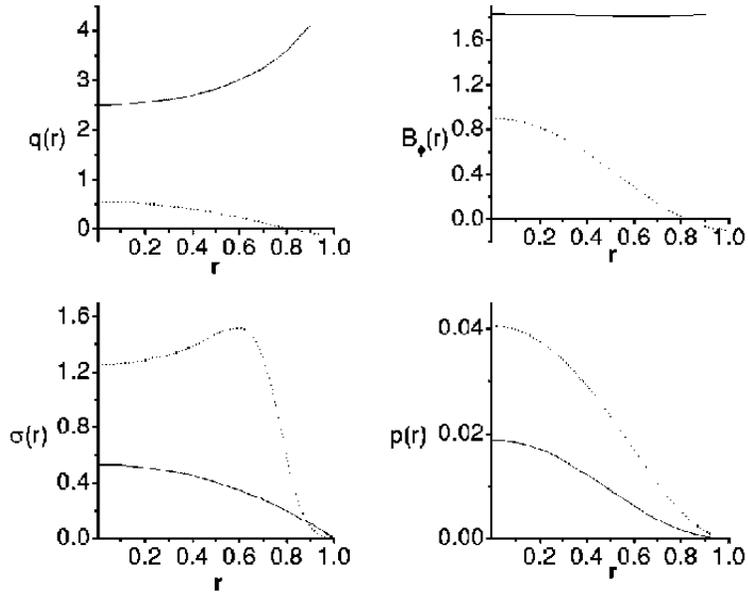, width=10.cm, height=8.cm}
\caption{\small\label{fig:toka-one} q-profile, $B_\phi, \sigma$,
and pressure profile of (i)Tokamak discharge (continuous line);
(ii) RFP discharge (dotted line); (\textit{from Bhattacharya et
al., 2000}).}
\end{figure}
\begin{figure}[ht!]
\epsfig{file=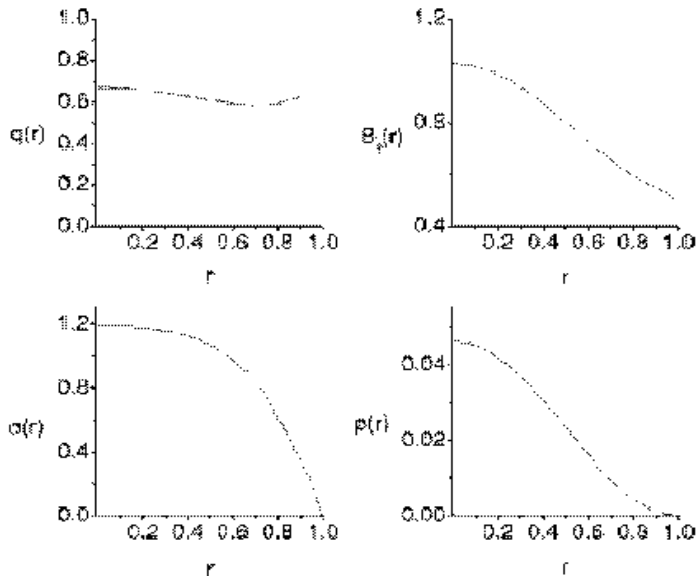, width=10.cm, height=8.cm}
\caption{\small\label{fig:toka-two} q-profile, $B_\phi, \sigma$,
and pressure profile of  low-q Discharge;  (\textit{from
Bhattacharya et al., 2000}).}
\end{figure}

A success of this theory is its applicability in a tokamak
configuration [Bhattacharya et al., 2000]. In toroidal geometry,
the CK eigenfunctions are hypergeometric functions in the large
aspect ratio approximation. Equilibria can be constructed under
the assumption that the total current J = 0 at the edge.  The
solution reproduces the q-profile, the toroidal magnetic field,
the pressure profile and $\sigma = {\bf J}\cdot {\bf B} /B^2$ for
tokamak, which are very close to the observed profiles (e.g.,
figure \ref{fig:toka-one}). The solutions allow for a tokamak,
low-q discharge as well as RFP like behavior with a change of
eigenvalue which would essentially determine the amount of
volt-sec associated with the discharge characterizing the relaxed
state. The poloidal Volt-sec/toroidal flux associated with the
discharge can be directly interpreted in terms of the parameter
$\lambda r_0$ ($r_0$ is the minor radius of the torus, $\lambda
=\sqrt[3]{\Lambda}$) and can be shown to increase with $\lambda
r_0$. MDR relaxed state can yield several kinds of solutions with
distinct q-profiles. For low values of $\lambda r_0$, the
q-profile is monotonically increasing and hence the solutions
resemble a tokamak type discarge together with a nearly constant
toroidal field, for larger values of $\lambda r_0$ the toroidal
field reverses at the edge, signifying an RFP-type behavior and
for intermediate values, the relaxed states indicate an ultra
low-q type of discharge exhibiting a q-profile with a pitch
minimum and $0<q<1 $. In figure \ref{fig:toka-one}, we plot the
toroidal magnetic field $B_\phi, \sigma = {\bf J}\cdot{\bf J}/B^2,
q$ and pressure profiles for different values of $\lambda r_0$,
for a fixed value of the aspect ratio = 4.0. The continuous line
corresponds corresponds to $\lambda r_0 = 1.5$, and shows a
monotonic increase with $r$, which is the essential feature of
tokamak discharge. For this case the the toroidal magnetic field
($B_\phi$) is essentially constant and the pressure profile has a
peak at the center of the minor cross section of the torus and
falls to zero at the edge. A non-constant behavior is shown by the
$\sigma$ profile. The dotted line corresponds to $\lambda r_0=
3.0$. In this case the toroidal field reverses sign at the edge,
which corresponds to an RFP like discharge. The q-profile shows a
similar behavior. The $\sigma$ profile has a dip at $r=0$ and
exhibits a bump near the outer region. Such profiles have been
experimentally observed in the ETA-BETA-II device [ Antoni et al,
1983]. For the values $\lambda r_0= 2.5$ the q-profile shows a
nonmonotonic behavior as shown in figure \ref{fig:toka-two}. The
toroidal magnetic field again peaks at the center and slowly
decreases towards the edge, showing a paramagnetic behavior that
is typical of ultra low-q discharge [Yoshida et al., 1986].
Pressure profile peaks at the center and goes to zero at the edge.
Some of these results have been observed in the SINP tokamak
[Lahiri et al, 1996] and were also reported earlier in REPUTE-1
[Yoshida et al., 1986] and in TORIUT-6 [Yamada et al, 1987].

This model has been utilized to obtain the spheromak configuration
from the spherically symmetric solution under appropriate boundary
conditions relevant to an insulating boundary at the edge
[Dasgupta et al., 2002]. These boundary conditions are are
consistent with the classical spheromak solution used by
Rosenbluth and Bussac [1979]. The most interesting part of this
spheromak solution  is the non-constant profile for ${\bf
J}\cdot{\bf B}/B^2$  plotted against normalized distance from the
magnetic axis.  This profile shows peaks outside the magnetic
axis, and this feature is qualitatively closer to the
experimentally observed profile reported by Hart et al. [Hart et
al., 1983].

The field-reversed configuration (FRC)  is a compact toroidal
confinement device [Tuszewski, 1988]  which is formed with
primarily poloidal magnetic field and  a zero or negligibly small
toroidal magnetic field. The plasma beta of the FRC is one of the
highest for any magnetically confined plasma.  FRC can be
considered as a relaxed state [Guo et al, 2005, 2006], but it is a
distinctly non-Taylor state. The FRC configuration can be obtained
for an externally driven dissipative system  which relaxes  with
minimum dissipation.  In Bhattacharya et al., [2001] we show that
the Euler-Lagrange equations for relaxation under MDR with
magnetic energy as the constraint represents an FRC configuration
for one choice of eigenvalues. This solution is characterized by
high beta, zero helicity and a completely null toroidal field and
supports  a non-constant pressure profile (non-force free).
Another choice of eigenvalue leads to a spheromak configuration.
A generalization of this work incorporating flow [Bhattacharya et
al., 2003] and taking magnetic and  flow energies and cross
helicities as constraints also produces the FRC configuration.
This state supports field aligned flows with strong shears  that
may lead to stability and better confinement.

\begin{figure}[ht!]
\epsfig{file=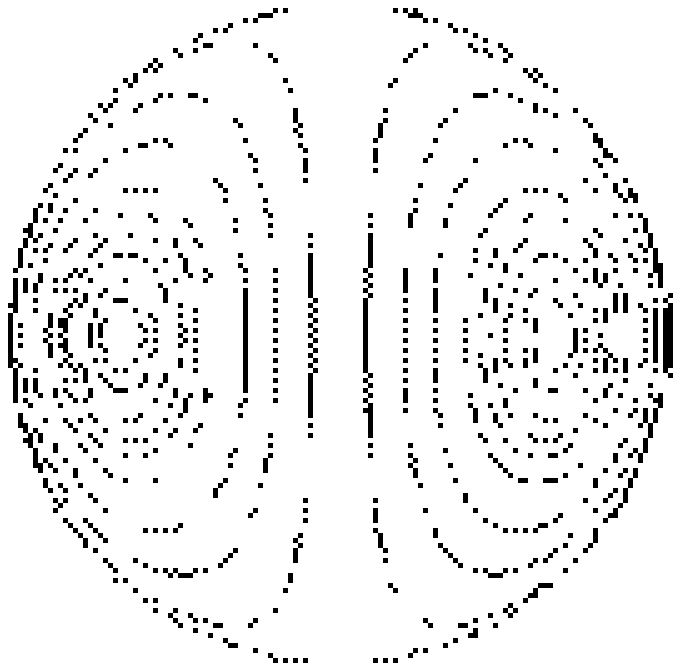, width=10.cm, height=8.cm}
\epsfig{file=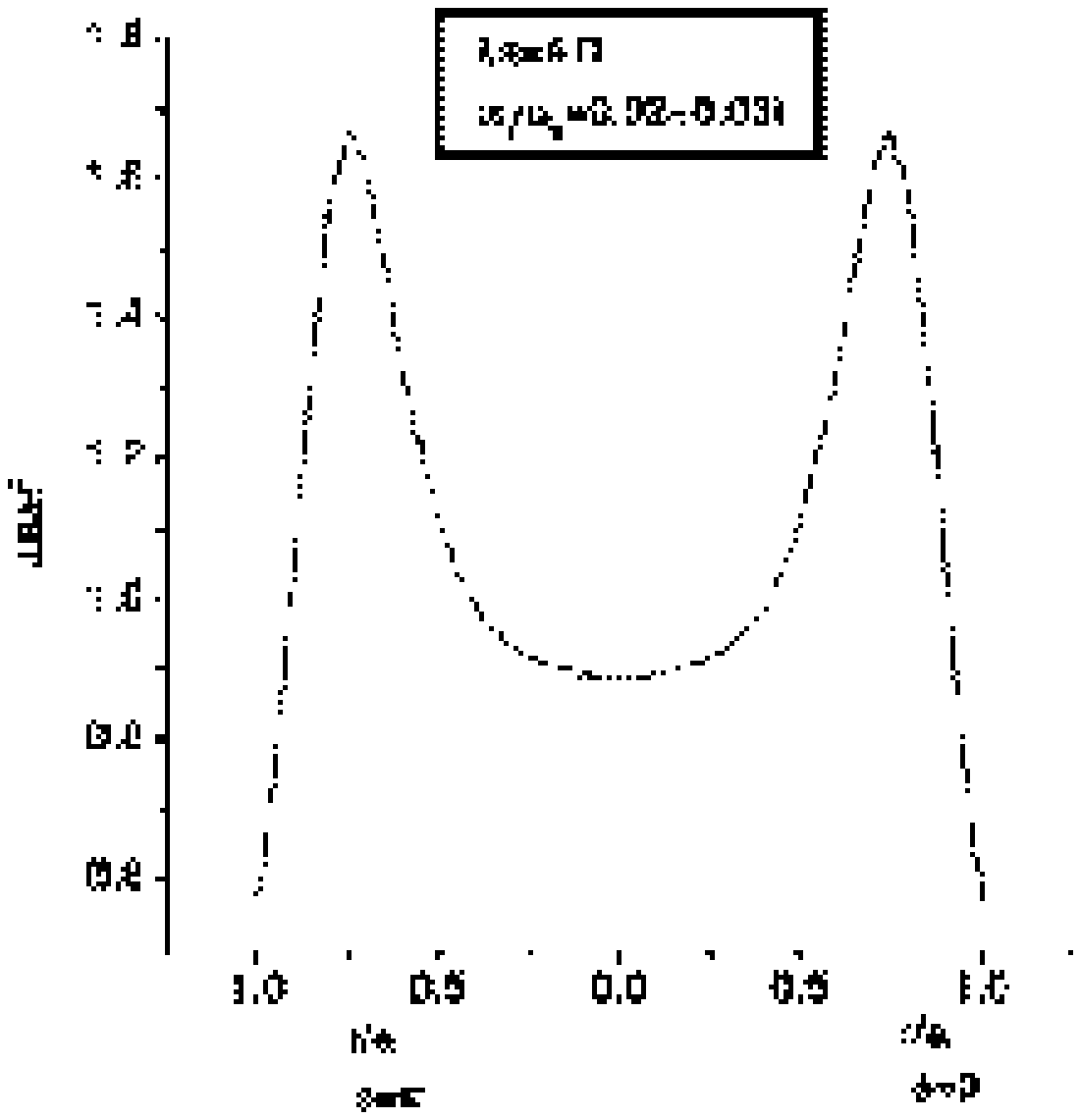, width=10.cm, height=8.cm}
\caption{\small\label{fig:sphero} Spheromak configuration (left
panel) and the non-constant $J_{||}/B$ derived from our model
based on MDR (right panel); (\textit{from Dasgupta et al.,
2002}).}
\end{figure}

\subsection{Simulation Results:} 
\begin{figure}[ht]
\begin{center}
\epsfig{file=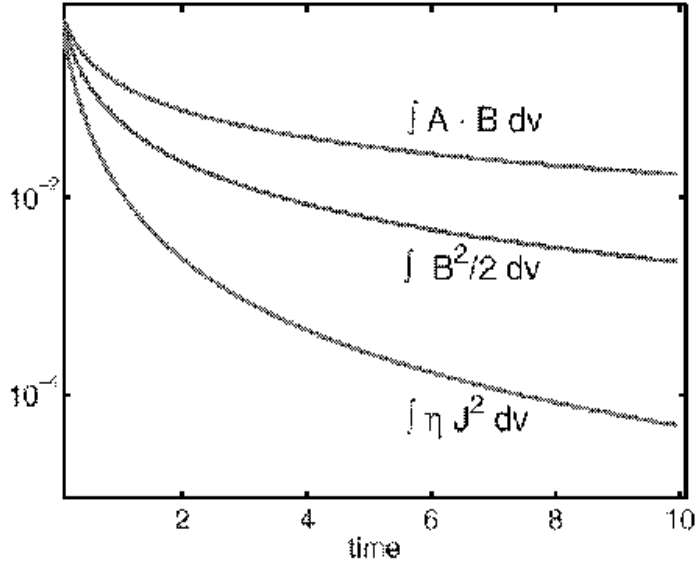, width=10.cm, height=8.cm}
\end{center}
\caption{\small\label{fig:sim-three} Time evolution of decay rates
associated with the turbulent relaxation of the ideal MHD
invariants, viz, global magnetic helicity $K_M= \int_V {\bf
A}\cdot{\bf B}dV$, magnetic energy $W= \int_V B^2/2dV$ and  the
total dissipation $R=\int_V \eta J^2 dV$, all shown in the same
scale. }
\end{figure}

To justify the  use of the MDR
principle in plasma relaxation and the choice of the total
dissipation rate as a minimizer in the variational principle we
performed a fully 3-D numerical simulations for a turbulent and
slightly dissipative plasma  using a fully compressible MHD code
[Shaikh et al., 2007] with periodic boundary conditions.
Fluctuations are Fourier expanded and nonlinear terms are
evaluated in real space. Spectral methods provide accurate
representation in Fourier space. In our code there is very little
numerical dissipation, and hence the ideal invariants are
preserved. The selective decay processes (in addition to
dissipation) depend critically on the cascade properties
associated with the rugged MHD invariants that eventually govern
the spectral transfer in the inertial range. This can be
elucidated as follows [Biscamp, 2003]. Magnetic vector potential
in 3D MHD dominates, over the magnetic field fluctuations, at the
smaller Fourier modes, which in turn leads to a domination of the
magnetic helicity invariant over the magnetic energy. On the other
hand, dissipation occurs predominantly at the higher Fourier modes
which give rise to a rapid damping of the energy dissipative
quantity $R$. A heuristic argument for this process can be
formulated in the following way. The decay rates of helicity $K$
and dissipation rate $R= \int_V\eta j^2 dV$ in the dimensionless
form, with the magnetic field Fourier decomposed as ${\bf B}({\bf
k},t)= \Sigma_k {\bf b_k}\exp(i{\bf k}\cdot{\bf r}) $, can be
expressed as,

\begin{equation}\label{comp1}
\frac{d K}{dt}= -\frac{2\eta}{S}\Sigma_k k {\bf b_k}^2;\qquad
\frac{d R}{dt}= -\frac{2\eta^2}{S^2}\Sigma_k k^4 {\bf b_k}^2
\end{equation}
where $S= \tau_R/\tau_A$, is the Lundquist number, $\tau_R,
\tau_A$ are the resistive and Alfv\'{e}n time scales,
respectively. The Lundquist number in our simulations varies
between $10^6$ and $10^7$. We find that at scale lengths for which
$k\approx S^{1/2}$, the decay rate of energy dissipation is $\sim
O(1)$ . But at these scale lengths, helicity dissipation is only
$\sim O(S^{-1/2})\ll 1$. This physical scenario is consistent with
our 3D simulations. Figure \ref{fig:sim-three} shows the time
evolution of the global helicity $K_M$, total magnetic energy
$W_M$ and the total dissipation rate $R$. It is seen that global
helicity remains approximately constant, (decaying very slightly
during the simulation time) while the magnetic energy decays
faster than the global helicity, but the energy dissipation rate
decays even faster than the magnetic energy.  The results of the
simulation show that total dissipation rate can serve as a
minimizer during a relaxation process.

\section{MDR based relaxation model for two fluid plasma with
external drive:} Recently, Bhattacharya and Janaki [2004]
developed a theory of dissipative relaxed states in two-fluid
plasma with external drive.  We describe the plasma (consisting of
ions and electrons) by the two-fluid equations. The basic
advantages of a two-fluid formalism over the single-fluid or MHD
counterpart are: (i) in a two-fluid formalism, one can incorporate
the coupling of plasma flow with the magnetic field in a natural
way, and (ii) the two-fluid description being more general than
MHD, is expected to capture certain results that are otherwise
unattainable from a single-fluid description. It should be
mentioned that although we are invoking the principle of minimum
dissipation rate of energy - instead of the principle of minimum
energy - our approach is closely similar to the approach to plasma
relaxation including Hall terms, as developed by Turner (1986).

In the two-fluid formalism the canonical momentum ${\bf{P}}_j$
canonical vorticity ${\bf{\Omega}}_j$, for the $j$-th species are
defined respectively as,

\begin{equation}
{\bf{P}}_j=m_j{\bf{u}}_j+\frac{q_j}{c}{\bf{A}}, \qquad
{\bf{\Omega}}_j=\nabla\times{\bf{P}}_j, \quad (j= i, e).
\end{equation}
A gauge invariant expression for the generalized helicity is
introduced by defining
\begin{equation}
K_j=\int_V{\bf{P}}_j\cdot{\bf{\Omega}}_j d\tau-
\int_V{\bf{P}}_j^\prime\cdot{\bf{\Omega}}_j^\prime dV;
\end{equation}
which is a relative helicity with respect to reference field
having ${\bf{P}}_j^\prime$  as the canonical momentum, and
${\bf{P}}_\alpha$ and ${\bf{P}}_\alpha^\prime$ are different
inside the volume of interest and the same outside. The gauge
invariance condition holds
\begin{equation}\label{omega}
({\bf{\Omega}}_j-{\bf{\Omega}}_j^\prime) \cdot\hat{n} =0
\Rightarrow ({\bf{B}}-{\bf{B}}^\prime)\cdot\hat{n}=0 \qquad
({\bf{\omega}}_j-{\bf{\omega}}_j^\prime) \cdot\hat{n}=0, \quad
({\bf{\omega}}_j=\nabla\times{\bf{u}}_j),
\end{equation}
under the condition that there is no flow-field coupling at the
surface. The above equations  then represent natural boundary
conditions inherent to the problem.

The helicity balance equation for the generalized helicity,
obtained from the balance equations for the generalized momentum
and vorticity, consists of two terms, the helicity injection terms
and helicity dissipation term (containing resistivity and
viscosity). The global helicity can be taken as a bounded function
following Montgomery  et al.,(1988) and it is possible to form a
variational problem  with the helicity dissipation rates as
constraints. Thus,
\begin{eqnarray}
\frac{d K_\alpha}{dt}&=&-\oint[G_\alpha-{\bf{P}}_\alpha\cdot{\bf
{u}}_\alpha]{\bf{\Omega}}_\alpha^\prime\cdot\hat{n}da-\oint({\bf{P}}_\alpha
\cdot{\bf{\Omega}}_\alpha){\bf{u}}_\alpha\cdot\hat{n}da \nonumber\\
&-&\oint\left(\frac{\partial{\bf{P}}_\alpha^\prime}{\partial t}
 \times{\bf{P}}_\alpha^\prime\right)\cdot\hat{n}da
 -\eta_\alpha^\prime\oint[\{\nabla\times{\bf{B}}+L_\alpha
 {\bf{\omega}}_\alpha)\}\times{\bf{P}}_\alpha]\cdot\hat{n}da \nonumber\\
 &-&2\eta_\alpha^\prime\int[\nabla\times({\bf{B}}+L_\alpha
 {\bf{\omega}}_\alpha)]\cdot{\bf{\Omega}}_\alpha d\tau
 \end{eqnarray}with
\begin{eqnarray}
 & &G_\alpha=\left[q_\alpha\phi+\frac{h_\alpha}{n_\alpha}-\frac{m_\alpha
 u_\alpha^2}{2}\right]; \qquad
\eta_\alpha^\prime=\frac{q_\alpha\eta c}{4\pi} \nonumber\\
 & & L_\alpha=\frac{4\pi}{q_\alpha c
 n_\alpha}\frac{\mu_\alpha}{\eta}\equiv~~{\rm{Prandtl~~number.}}
\end{eqnarray}The minimization integral is then obtained as,
\begin{eqnarray}
 {\cal{I}}&=&\int
 \left[(\nabla\times{\bf{B}})\cdot(\nabla\times{\bf{B}})+\frac{4\pi
 n q L_i}{c}{\bf{\omega}}_i\cdot{\bf{\omega}}_i\right]d\tau \nonumber\\
 &+&\lambda_i\int[\nabla\times({\bf{B}}+L_i{\bf{\omega}}_i)]\cdot{\bf{\Omega_i}}
 d\tau-\lambda_e\int(\nabla\times{\bf{B}})\cdot{\bf{\Omega}}_e
 d\tau
 \end{eqnarray}

\noindent where $L_i = 4\pi\mu_i/ecn_i \eta$ is the magnetic
Prandtl number for ions, $\mu_i$ is the ion viscosity and $\eta$
is the resistivity. We assumed the electron mass $m_e\rightarrow
0. $ The first two terms in the above integral represent the ohmic
and viscous dissipation rates while the second two integrals
represent the ion and electron generalized helicity dissipation
rates. $\lambda_i$ and $\lambda_e$ are the corresponding Lagrange
undetermined multipliers.The Euler-Lagrange equations are written
as,\begin{eqnarray} &
&\nabla\times\nabla\times{\bf{B}}+\frac{(\lambda_i+\lambda_e)q}{c}
\nabla\times){\bf{B}}+\frac{\lambda_i}{2}\left( \frac{q
L_i}{c}+m_i\right)(\nabla\times){\bf{\omega}}_i=\nabla\psi;
\label{mageq}\\
& &\nabla\times\nabla\times{\bf{\omega}}_i+\frac{2\pi n
q}{m_i\lambda_i c}\nabla\times{\bf{\omega}}_i+\frac{1}{2 m_i
L_i}\left( \frac{q L_i}{c}+m_i\right)(\nabla\times)^2{\bf{B}}=
\nabla\chi;\label{floweq}
\end{eqnarray}
where $\psi$ and $\chi $ are arbitrary gauge functions, with
$\nabla^2\psi=\nabla^\chi=0 $. Following Turner (1986) we
can write the exact solution of the coupled equations
(\ref{mageq}) and (\ref{floweq}) as a linear superposition of two
CK eigenfunctions, which are the solutions of
$\nabla\times{\bf{Y}}_k=\lambda_k{\bf{Y}}_k $ i.e.,

\begin{figure}[ht]
\begin{center}
\epsfig{file=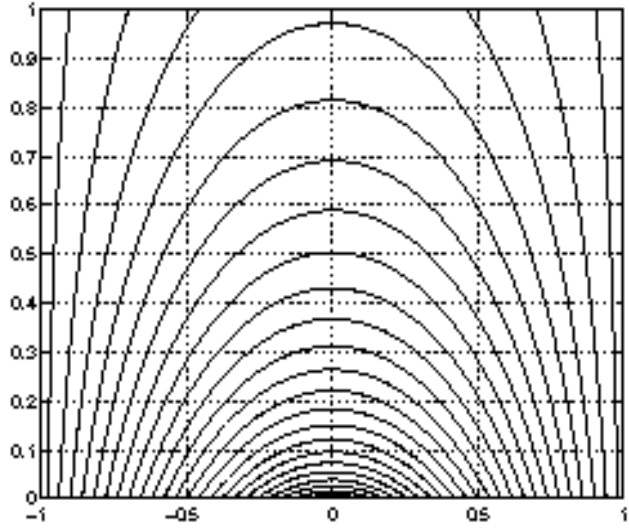, width=10.cm, height=8.cm}
\end{center}
\caption{\small\label{fig:flow} Flow lines in the $x$-$z$ plane
for $\gamma=1.5$ (\textit{from Bhattacharya et al., 2007}).}
\end{figure}

\begin{equation}
{\bf{B}}={\bf{Y}}_1+\alpha{\bf{Y}}_2 \qquad
{\bf{\omega}}_i={\bf{Y}}_1+\beta{\bf{Y}}_2,
\end{equation}
and $\alpha$ and $\beta$ are constants and quantify the non
force-free part, to  be obtained from some boundary conditions.

\subsection{Application to solar arcade problem}

One of the important features  of this formalism is the presence
of a nontrivial  flow and the coupling between the  magnetic field
and the  flow. The solutions of the corresponding Euler-Lagrange
equation highlight the role of nonzero plasma flow in securing a
steady MDR relaxed state  for an open and externally driven
system.  This model has been very successful in describing the
arcade structures observed in Solar corona [Bhattacharya et al.,
2007]. To model solar arcade-type magnetic fields, we make some
simplified assumptions and consider a 2-D magnetic field in the
$x$-$z$ plane with the $x$ direction lying on the photosphere
surface and $z$ is the vertical direction. With appropriate
boundary conditions and after some lengthy algebra, the solutions
for the components of the magnetic fields are obtained as,

\begin{figure}[ht]
\begin{center}
\epsfig{file=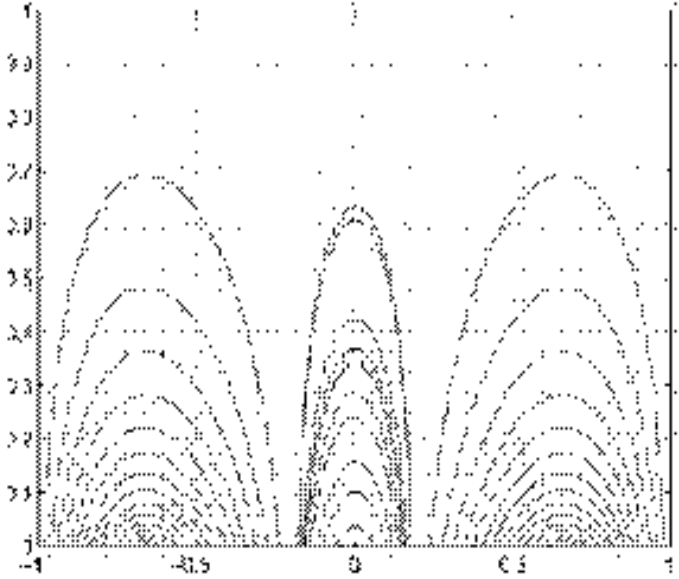, width=10.cm, height=8.cm}
\end{center}
\caption{\small\label{fig:arcade}Magnetic field lines in the
$x$-$z$ plane corresponding to (i) single arcade structure for
$\gamma=1.5$ in left panel and (ii) triple arcade structure for
$\gamma=3.36$ in right panel (\textit{from Bhattacharya et al.,
2007}).}
\end{figure}

\begin{eqnarray}
& &B_x=\sqrt{\gamma^2-1}\left[2\cos\gamma
x-\frac{2\cos\gamma}{\cos\sqrt{\gamma^2-1}}\cos(x\sqrt{\gamma^2-1})\right]
e^{-z\sqrt{\gamma^2-1}};\nonumber\\
& &B_y=0;\\
& &B_z=-\left[2\gamma\sin\gamma
x-\frac{2\sqrt{\gamma^2-1}\cos\gamma}{\sin\sqrt{\gamma^2-1}}\sin(x\sqrt{\gamma^2-1})\right]
e^{-z\sqrt{\gamma^2-1}}\nonumber;
\end{eqnarray}
with $\gamma = \sqrt{(k^2+\lambda^2)/\lambda^2}$ ($\lambda$ is the
eigenvalue and $k$ is the  wave number associated with the
solution). The same procedures can be adopted for the flow
pattern. In figure \ref{fig:arcade}  the magnetic field lines in
the$x$-$z$ plane is plotted for two different values of the
parameter $\gamma$. The arcade structures are obtained by solving
the field line equations $\frac{dz}{dx}=\frac{B_z}{B_x}$.  It is
interesting to note that the arcade structure changes with the
change of the single parameter $\gamma$. Figure \ref{fig:flow}
plots the flow lines for $\gamma = 1.5$. In cylindrical geometry
the solutions of eqn. (\ref{mageq}) can be written in terms of a
superposition of Bessel functions. Such solutions yield 3-D Flux
ropes, which are plotted in the Figure
({\ref{fig:fluxrope}).

\begin{figure}[ht]
\begin{center}
\epsfig{file=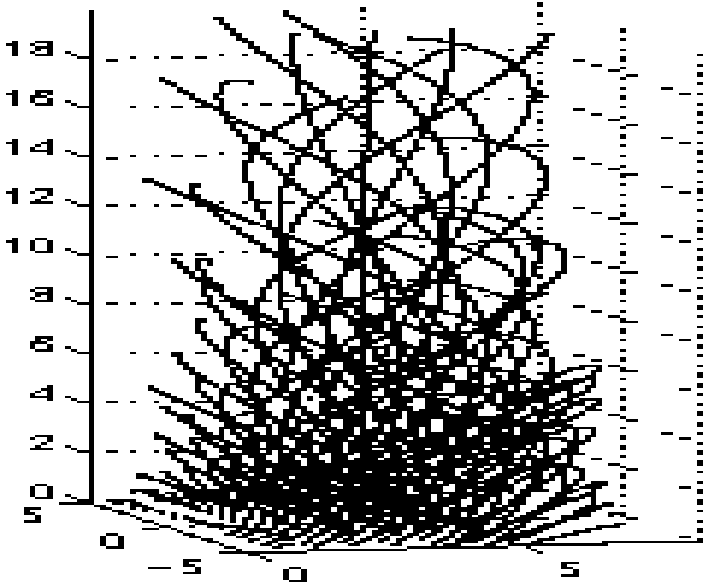, width=10.cm, height=8.cm}
\end{center}
\caption{\small\label{fig:fluxrope} Plot of 3-D Fluxrope obtained
from the solution eqn.(\ref{mageq}).}
\end{figure}

\subsection{Non-force Free Coronal Magnetic Field Extrapolation
Based on MDR:}  An outstanding problem in solar physics is to
derive the coronal magnetic field structure from measured
photospheric and/or chromospheric magnetograms. A new approach to
deriving three-dimensional non-force free coronal magnetic field
configurations from vector magnetograms, based on the MDR
principle has been recently developed [Hu, Dasgupta, \& Choudhary
(2007)]. In contrast to the principle of minimum energy, which
yields a linear force-free magnetic field, the MDR gives a more
general non-force free magnetic field with flow. The full
MDR-based approach requires two layers of vector magnetic field
measurements. Its exact solution can be expressed as the
superposition of two linear force-free fields with distinct
$\lambda$ parameters, and one potential field. The final solution
is thus decomposed into three linear force-free extrapolations,
with bottom boundary conditions derived from the measured vector
magnetograms, at both photospheric and chromospheric levels. The
semi-analytic test case shown in Figure~\ref{FIGHu} illustrates
the feasibility and the high performance of this method. This
extrapolation is easy to implement and much faster than most other
nonlinear force-free methods [Hu \& Dasgupta 2007]. It also takes
full advantage of multiple layer solar magnetic field
measurements, thus gives a more realistic result.

\begin{figure}[ht]
\begin{center}
\epsfig{file=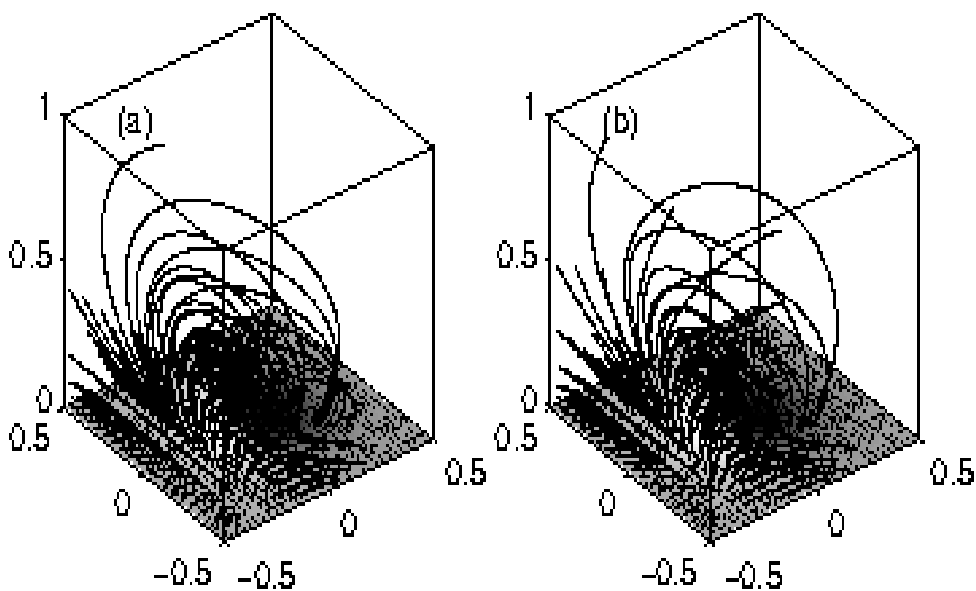, width=13.cm, height=8.cm}
\end{center}
\caption{3D view of magnetic field lines in a box simulating the
typical extrapolation domain. (a) exact solution; (b) numerical
extrapolation result. Bottom image at height 0 shows the normal
field component distribution with gray scale indicating strongly
negative (black) to positive (white) field. The great ressemblance
between the exact and the numerical solutions justifies the
feasibility and high performance of this method (Hu, Dasgupta, \&
Choudhary 2007).}
\label{FIGHu}
\end{figure}

\section{Conclusion}

We have demonstrated that a relaxation model based on the minimum
dissipation of energy can successfully explain some of the salient
observations both in laboratory and space plasma. An MDR based
model model in its simpler form is capable of yielding basic
characteristics of most of the laboratory plasma confinement
devices, and a generalization of this model for two fluid with
external drive can couple magnetic field and flow and predict
solar arcade structures. The above findings definitely points out
that the MDR relaxed states applied to astrophysical plasmas is a
worthy case for further investigations. In future we plan to apply
this model to investigate magnetic field structure in the sun with
an aim to understanding coronal mass ejection.

\section{Acknowledgement}

We gratefully acknowledge partial supports from NASA LWS grant
NNX07A073G, and NASA grants NNG04GF83G, NNG05GH38G, NNG05GM62G, a
Cluster University of Delaware subcontract BART372159/SG, and NSF
grants ATM0317509, and ATM0428880.

\label{lastpage}
\end{document}